# Wear Analysis of a Heterogeneous Annular Cylinder

Fabian Forsbach, Maximilian Schuster, Daniel Pielsticker, Qiang Li[*]
Technical University of Berlin, Berlin, Germany
Corresponding author: Qiang Li, qiang.li@tu-berlin.de

**Abstract**

The wear of a cylindrical punch composed by two different materials alternatively distributed in annular forms is studied with the Method of Dimensionality Reduction (MDR). The changes in surface topography and pressure distribution during the wear process is obtained and validated by the Boundary Element Method (BEM). The pressure in each annular ring approaches constant in a stationary state where the surface topography does not change any more. Furthermore, in an easier way, using the direct integration the limiting profile in steady wear state is theoretically calculated, as well as the root mean square (RMS) of its surface gradient which is closely related to the coefficient of friction between this kind of surface and an elastomer. The dependence on the wear coefficients and the width of the annular areas of two phases is obtained.

## 1. Introduction

The friction between a vehicle's tire and the road is an everyday contact problem. Due to the movement, the road is polished and its resistance against skid of the contact is decreased. To improve the quality and durability of the contact between road and tire, knowledge about the time-dependent behavior of the contact is vital. Recent research [1][2] covers an experimental approach to this contact problem to determine the long-time behavior of different construction materials being exposed to the polishing effects of the tires of vehicles. The behavior of the road material's skid resistance is examined. Paper [1] presents the dominating effect of the aggregates on friction, as soon as the binder is polished after a given time of usage, which is explained by a modification of the aggregate's micro texture. [2] examines the relation between the composition of the asphalt and the friction that it provides in the tire to road contact. It results in the introduction of a hardness parameter related to the friction coefficient.

In [3] it is shown that the friction coefficient is approximately the mean slope of the surface for the contact between an elastomer and a rigid rough surface. In this paper we investigate the wear of a multi-phase cylinder in sliding contact with an elastic half space. In detail, the limiting profile and its surface's gradient are determined. Under the assumption of Archard's law of wear [4], stating that the volume of the material loss $\Delta V$ is proportional to the normal load $F_N$ and sliding distance $x$, and inversely proportional to the hardness of the material $\sigma_0$,

$$\Delta V = k_{wear} \frac{F_N x}{\sigma_0}, \qquad (1)$$

the change of the surface topography of the cylinder during the sliding is calculated numerically using the Method of Dimensionality Reduction (MDR). The MDR was proposed by Popov and Heß for the fast calculation of various contact problems [5]. It provides exact solutions for axisymmetric contacts by mapping of three-dimensional contacts onto one-dimensional contacts, and has been applied to studying on different wear problems, e.g. fretting wear [6] and gross slip wear [7] which were also validated by experiment [8] or by the Finite Element Method [9]. Therefore, we introduce a heterogeneous annular cylinder with rings of alternating material representing aggregate and binder respectively to study the effects of material composition. Furthermore, to study the surface gradient of the contact body in stationary state, a direct integration is applied to calculate the final surface topography quickly.

The parts of this paper are organized as following. Section 2 describes the model of heterogeneous cylinder. Section 3 presents the simulation of wear process from the first contact to the final stationary state using the MDR. Section 4 gives a method for direct calculation of the profile in stationary state and discusses the influence of ratio of wear coefficients of two materials and their area ratio on the surface gradient. At last a short conclusion is presented in section 5.

## 2. Heterogeneous cylinder

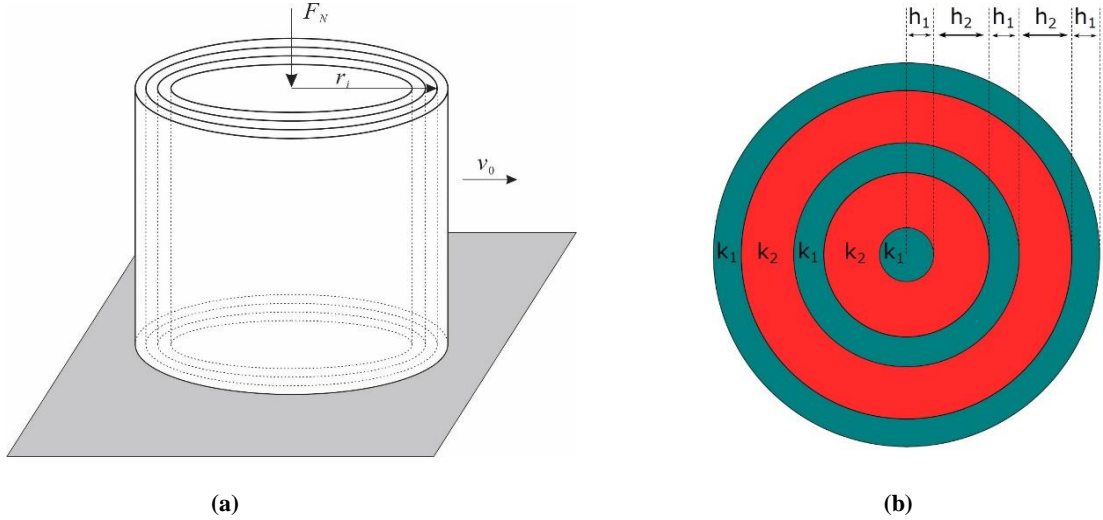

(a)          (b)

Fig. 1 (a) The cylinder made of a number of rings with radius $r_i$. (b) Exemplary surface with wear coefficients $k_{1,2}$ and ring widths $h_{1,2}$ for a cylinder with five rings.

We consider the contact as follows. As shown in Fig. 1a, the cylinder is pressed into the elastic half space under the load $F_N$, and moves tangentially with a constant velocity $v_0$. The initially flat cylinder surface will thus be worn and approaches a limiting profile over time that is to be found. The cylinder is structured with the contact radius $a$ composed of a number of rings $N_R$ with the radius $r_i = ia/N_R$. The two different materials are regarded by two different wear coefficients $k_{1,2}$ alternating with each ring. Furthermore, the widths $h_{1,2}$ of the rings can be variable, as shown in Fig. 1b. In the case of a tire-asphalt contact, these materials could correspond to the rock particles and binder of the asphalt.

## 3. Calculation of wear process using the MDR

### 3.1 Method of Dimensionality Reduction

The Method of Dimensionality Reduction gives exact solution of axially symmetric contact problems [5][10], including the case of this annular cylinder with rings. In this study it will be used to quickly determine the stress distribution of a contact.

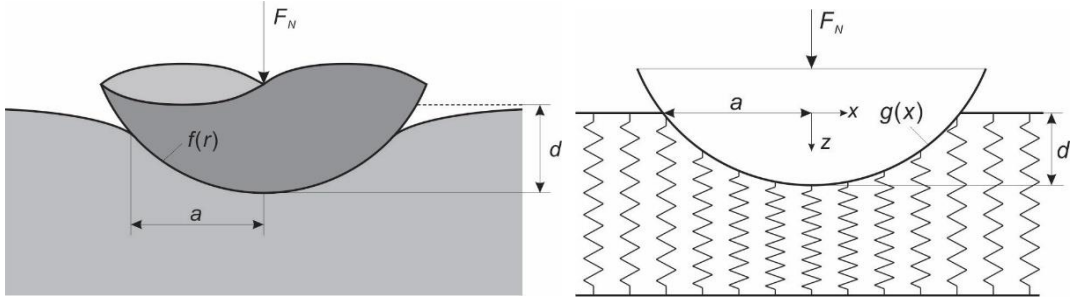

**Fig. 2 3D contact (left) and its equivalent one-dimensional model (right).**

The idea of MDR is to reduce the dimensionality of the contacting bodies, meaning that instead of three-dimensional elastic bodies, a one-dimensional linearly elastic foundation is used for the contact. The foundation is handled as a linear array of spring elements with an equal distance $\Delta x$. Fig. 2 shows the array of springs in contact with an arbitrary rigid body. In a normal contact of a rigid body and an elastic half space the stiffness of the springs is only considered in z-direction with $E^*$ being the effective elastic modulus:

$$\Delta k_z = E^* \Delta x \text{ , with } E^* = \frac{E}{1-\nu^2}, \tag{2}$$

where $E$ and $\nu$ are the elastic modulus and Poisson's ratio of the elastic half space. To use this one-dimensional bedding, it is necessary to transform the three-dimensional contact profile $f(r)$ into the one-dimensional MDR-profile $g(x)$ by the following transformation

$$g(x) = |x| \int_0^{|x|} \frac{f'(r)}{\sqrt{x^2 - r^2}} dr . \tag{3}$$

This profile can be used to calculate the relevant parameters of the contact problem. The displacement in z-direction can be found by the following equation with $d$ being the indentation depth:

$$u_z(x) = d - g(x) . \tag{4}$$

The proportionality of displacement $u_z(x)$ and force at this point $\Delta F_z(x)$ can be used to define the linear force density $q_z(x)$:

$$\Delta F_z(x) = \Delta k_z u_z(x) = E^* u_z(x) \Delta x \tag{5}$$

$$q_z(x) = \frac{\Delta F_z(x)}{\Delta x} = E^* u_z(x) = E^*\left[d - g(x)\right] \tag{6}$$

Finally, this linear force density is used to calculate the stress distribution $p(r)$

$$p(r) = -\frac{1}{\pi}\int_r^\infty \frac{q'(x)}{\sqrt{x^2 - r^2}} dx \tag{7}$$

## 3.2 Simulation procedure

The algorithm employed to solve the contact problem with the method of dimensionality reduction (MDR) is outlined in the flowchart in Fig. 3.

Single steps will be explained in the following. Steps (2)-(6) are repeated until the profile reaches the limiting profile within a given tolerance.

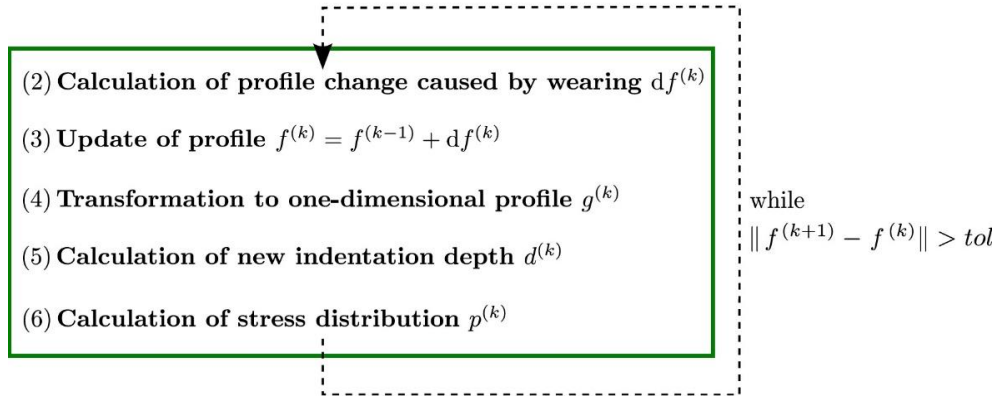

Fig. 3 Solving algorithm

### (a) Initial values (0)-(1)

The initial cylindrical profile $f(r)$ and the respective one-dimensional MDR-profile $g(x)$ are given by

$$f^{(0)}(r) = \begin{cases} 0 & \text{for } r < a \\ \infty & \text{for } r > a \end{cases}, \tag{8}$$

$$g^{(0)}(x) = \begin{cases} 0 & \text{for } x < a \\ \infty & \text{for } x > a \end{cases}. \tag{9}$$

The initial indentation depth under a constant normal force $F_N$ can be estimated according to [3]:

$$d^{(0)} = \frac{F_N}{2aE^*}. \tag{10}$$

The initial pressure distribution is

$$p^{(0)} = p_0 \left(1 - \frac{r^2}{a^2}\right)^{-1/2} \quad \text{with} \quad p_0 = \frac{F_N}{2\pi a^2}. \tag{11}$$

**(b) Change of profile due to wear (2)-(3)**

The heterogeneous, cylindrical indenter with profile $f(r)$ is pressed into an elastic half space with a constant normal force $F_N$ and then moved sideways with a constant velocity $v_0$. Archard's law of wear (1) for this heterogeneous indenter can be written as:

$$\Delta V = k_{wear}(r) \frac{F_N x}{\sigma_0(r)} \tag{12}$$

where $k_{wear}(r)$ and $\sigma_0(r)$ are the radius-dependent wear coefficient and hardness. Defining a new wear coefficient $k(r) = k_{wear}(r)/\sigma_0(r)$, considering the relative movement and the fact that for a cylinder the contact area is constant yields

$$df(r) = k(r) p(r) dx = k(r) p(r) v_0 dt \tag{13}$$

where $p(r)$ is the normal stress. This equation is used to calculate the profile change for a time step $dt$. Thus the profile change of the $k$-th time step is

$$df^{(k)}(r) = k(r) p^{(k-1)}(r) v_0 dt \tag{14}$$

with a piecewise constant distribution of wearing coefficients as depicted in Fig. 1. The profile is updated by

$$f^{(k)}(r) = f^{(k-1)}(r) + df^{(k)}(r). \tag{15}$$

**(c) Stress distribution of the updated profile (4)-(6)**

In order to obtain the profile change in Eq. (14), the new one-dimensional profile $g^{(k)}(x)$ and the new indentation depth $d^{(k)}$ are needed. The former one can be calculated by transformation of the updated profile $f^{(k)}(r)$ with Eq. (3), and the latter can be found by using the equilibrium of constant normal force and linear force density $q_z(x)$ (see Eq. (6))

$$F_N = \int_{-a}^{a} q_z(x) dx = 2E^* d^{(k)} a - 2E^* \int_{0}^{a} g^{(k)}(x) dx \tag{16}$$

Thus the new indentation depth is

$$d^{(k)} = \frac{F_N}{2E^*a} - \frac{1}{a}\int_0^a g^{(k)}(x)\,\mathrm{d}x \tag{17}$$

Now the pressure distribution $p^{(k)}(r)$ can be calculated by integrating Eq. (7) numerically.

### 3.3 Numerical results

Fig. 4 shows an example of the time-dependence of wear of the indented cylinder with five rings as described in Fig. 1b: 3D profiles $f(r)$, and corresponding 1D profile $g(x)$ as well as stress distributions $p(r)$ at distinct time steps. The analytical solution for an unworn cylinder is displayed by a dotted line for comparison. The profiles are normalized with the initial indentation depth $d_0$ calculated as in Eq.(10), the stress is normalized by $p_0$ as in Eq.(11) and $r$ and $x$ respectively by the radius of the cylinder $a$. Furthermore, the numerical results from the simulation with the boundary element method [11] are added in this figure for comparison. It can be seen that the MDR the BEM match within a small tolerance.

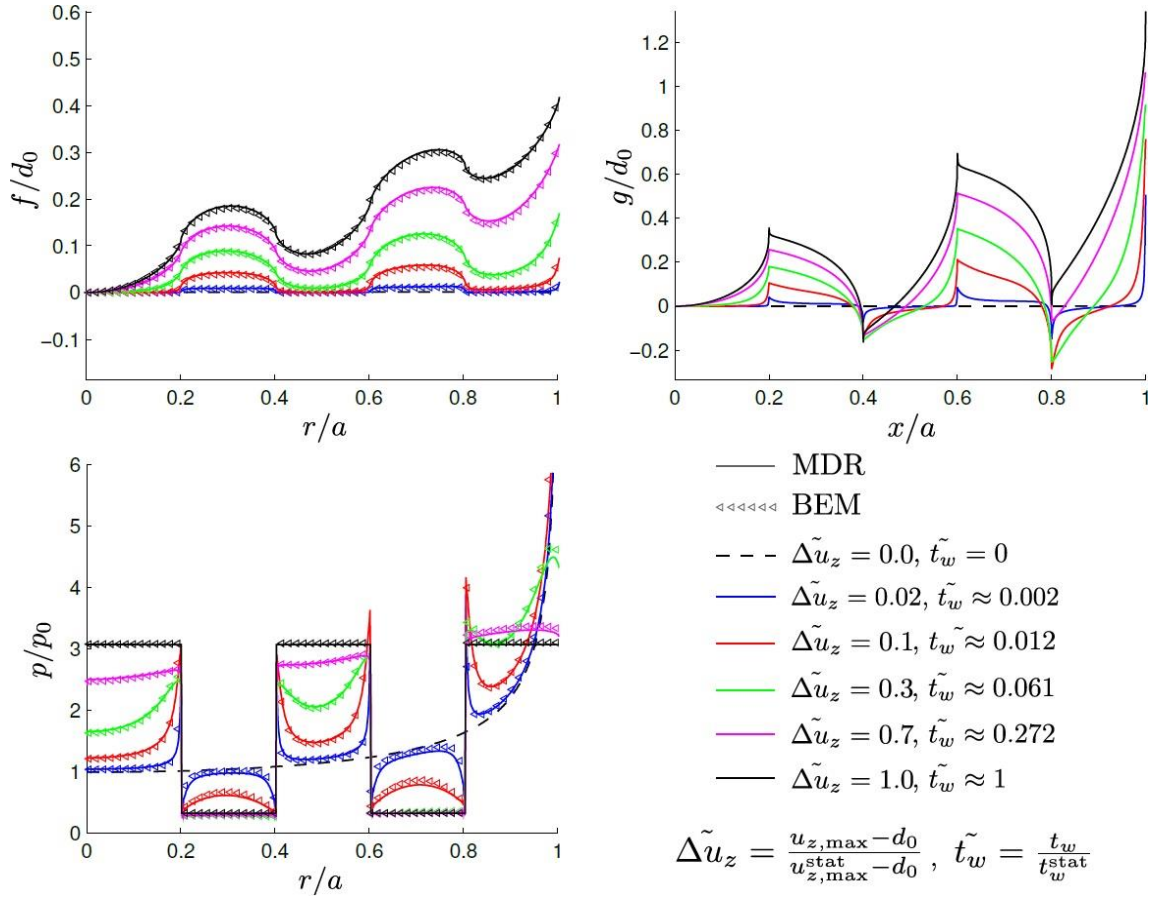

**Fig. 4 Worn profiles and pressure distribution during the wear process for the cylinder with 5 rings, $k_2/k_1$=10 and $h_2/h_1$=1.**

Two parameters are monitored in each time step: Firstly, the difference of maximal displacement of the elastic half-space $u_{z,\max}$ and initial indentation depth $d_0$ and secondly the elapsed time $t_w$ which is equivalent to the sliding distance due to constant sliding velocity. Both parameters are normalized by the according values of the limiting profile $u_{z,\max}^{stat}$ and $t_w^{stat}$.

Obviously, wear occurs on all parts of the cylinder as Eq.(13) indicates. The profile however changes as long as the product $C(r) = p(r)k(r)$ is not constant. Thus, for the piecewise constant distribution of wear coefficients the resulting stress distribution of the limiting profile must be piecewise constant as well with $p_1 > p_2$ in this case ($k_1 < k_2$). This can be observed for the limiting profile.

In the investigated case the stress level of rings with the lower wear coefficient is more than three times greater than $p_0$ and less than half for rings with the higher coefficient. A closer look at the time steps shows that the profile changes very quickly close to the edge of the body where the stress is theoretically infinite in the beginning. Further it changes quicker in regions with higher wear coefficient (in this case $k_2$). This profile change results in a reduction of stress in these regions. Due to the constant normal force and considering the equilibrium of forces this also results in a rise of stress in regions that are not subjected to a lot of profile change initially. As a consequence, these regions are in the following also subjected to profile change.

The constant stress level of rings with the higher wear coefficient (in this case $k_2$ and $p_2$) is reached significantly faster. Therefore, the time needed to reach the limiting profile is mostly dependent on the lowest wear coefficient.

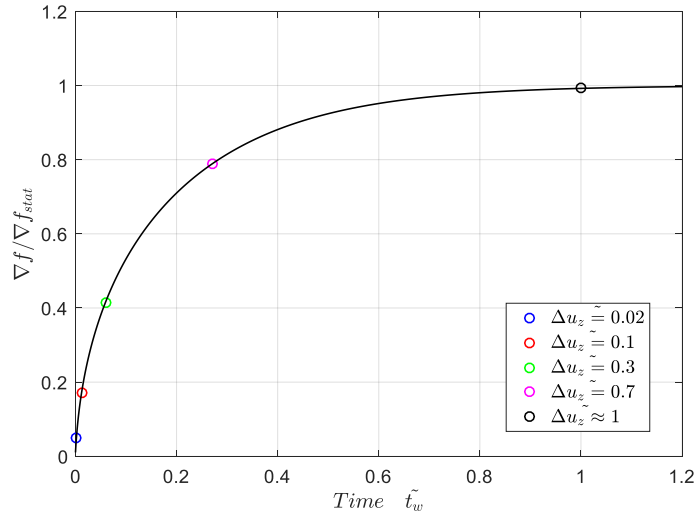

**Fig. 5 Change of normalized RMS of the surface gradient during the wear process.**

Looking at the monitored parameters $\Delta \tilde{u}_z$ and $\tilde{t}_z$ we observe that the profile change slows down significantly as the stress differences within the rings are getting smaller. For 70% of displacement difference only 27% of the time needed to reach the limiting profile has elapsed. This behavior can also be observed in Fig. 5 which shows the change of the surface gradient of the cylinder, where the cycles indicates the states at the time moment of the Fig. 4. The root mean square (RMS) of the surface gradient is defined as

$$\nabla f = \sqrt{\left\langle \left( \frac{\mathrm{d} f_{stat}(r)}{\mathrm{d} r} \right)^2 \right\rangle}. \qquad (18)$$

## 4. Limiting Profile and surface gradient

### 4.1 Calculation of limiting profile by direct integration

As another approach to quickly determine the limiting profile, this method of direct integration is used as well. As shown in [3], the surface displacement $u_z$ of an elastic half space at a certain point $r$ generated by a thin ring is given by

$$u_z(r) = \frac{F_N}{2\tilde{r}E^*} \frac{4}{\pi^2(1+r/\tilde{r})} K\left(2\frac{\sqrt{r/\tilde{r}}}{1+r/\tilde{r}}\right). \tag{19}$$

Therein $\tilde{r}$ is the radius of the ring and $K$ is the elliptic integral

$$K(\kappa) = \int_0^{\frac{\pi}{2}} \frac{d\varphi}{\sqrt{1-\kappa^2 \sin^2\varphi}}. \tag{20}$$

In order to find the surface displacement at point $r$ resulting from an entire cylinder rather than just a thin ring Eq. (19) is integrated

$$u_{z,tot}(r) = \frac{4}{\pi E^*} \int_0^a K\left(2\frac{\sqrt{r/\tilde{r}}}{1+r/\tilde{r}}\right) \frac{p(\tilde{r})}{1+r/\tilde{r}} d\tilde{r} \tag{21}$$

with the normal force $F_N$ being

$$F_N = 2\pi \int_0^a \tilde{r} p(\tilde{r}) d\tilde{r}. \tag{22}$$

The shape of the given cylinder alters until a steady wear, and thus a profile change

$$\frac{df(r)}{dt} = k_{1,2}(r) p(r) v_0 \equiv \text{const} \tag{23}$$

is reached. Consequently, for the limiting profile, the stress within each ring needs be constant:

$$p_{stat}(r) = \begin{cases} p_1 & \text{for } r \in \mathbb{k}_1 = \{r | k(r) = k_1\} \\ p_2 = k_1/k_2 \cdot p_1 & \text{for } r \in \mathbb{k}_2 = \{r | k(r) = k_2\} \end{cases}. \tag{24}$$

Here $\mathbb{k}_1$ and $\mathbb{k}_2$ are the sets of $r$ corresponding to odd respectively even rings. By inserting Eq. (24) in the equation for the normal force

$$F_N = 2\pi \int_0^a \int_0^{2\pi} p_{stat}(r) r dr \tag{25}$$

the stress in rings with odd numbers can be calculated to

$$p_1 = \frac{F_N}{2\pi}\left(\int_{\mathbb{k}_1} r\mathrm{d}r + \frac{k_1}{k_2}\int_{\mathbb{k}_2} r\mathrm{d}r\right)^{-1}. \tag{26}$$

This stress distribution can now be inserted in Eq. (21) to find the stationary solution for the surface displacement $u_{z,tot}(r)$. Finally, the limiting profile of the indenter is given by

$$f_{stat}(r) = u_{max} - u_{z,tot}(r), \tag{27}$$

where $u_{max}$ is the maximum value of $u_{z,tot}(r)$.

## 4.2 Results

With the direct integration method described above, the limiting profiles as well as their surface gradient for a variety of parameters $k_2/k_1$ and $h_2/h_1$ were determined. Dependence of friction on wear coefficients and ring widths are shown in Fig. 6 and Fig. 7. The surface gradient normalized by $\nabla f_0$ - the case of a homogeneous cylinder ($k_2/k_1 = 1$) is plotted versus the ratio of wear coefficients $k_2/k_1$ and ring widths $h_2/h_1$ respectively. Each curve represents the number of rings used as it was introduced in Fig. 1.

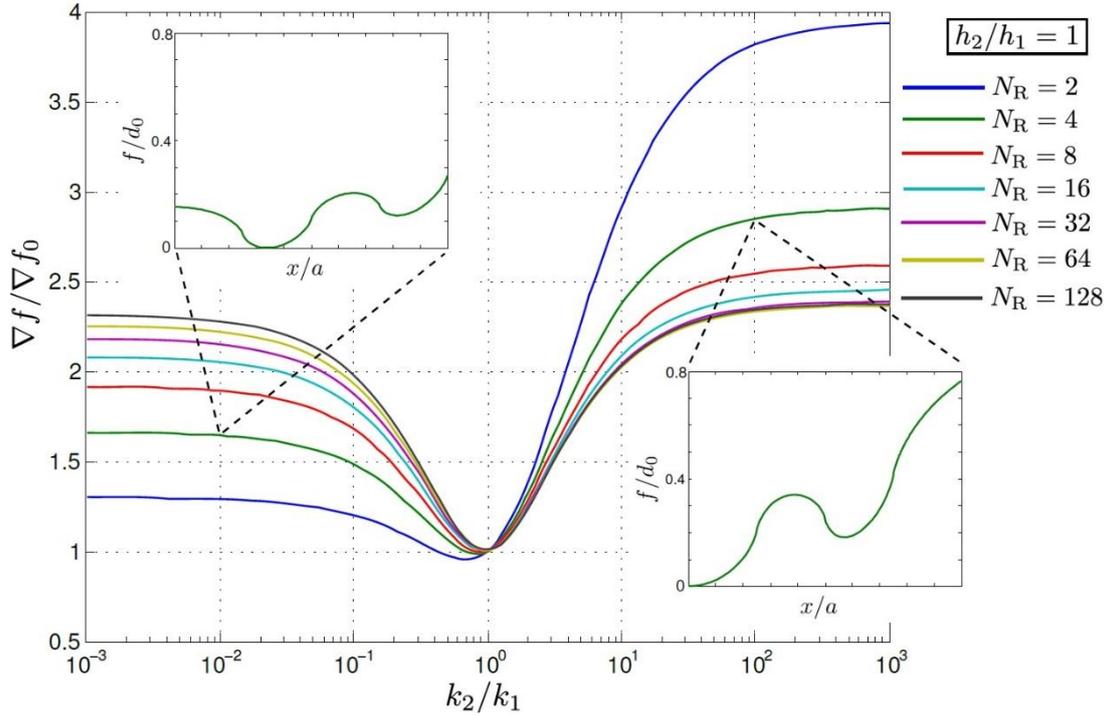

**Fig. 6 Normalized RMS of the surface gradient as a function of $k_2/k_1$ for different numbers of rings. The small figures show limiting profiles for extracted configurations.**

For a fixed ratio of ring widths $h_2/h_1 = 1$ and variable ratio of wear coefficients $k_2/k_1$ (Fig. 6) the plots are symmetrical for large numbers of rings. In all cases the surface gradients are converging towards different values for $k_2/k_1 > 100$ and $k_2/k_1 < 0.01$. This means for very large or very small ratios the influence of the faster wearing rings can be neglected and the problem is equivalent to single rings of the slower wearing material. The limiting value for large numbers of rings is $\nabla f/\nabla f_0 \approx 2.4$. For ratios in the range $0.01 < k_2/k_1 < 100$ the surface gradient is obviously very susceptible to the materials chosen.

In the case of very few rings the plots are asymmetrical and depend on the order of materials within the cylinder. This can be observed in the smaller figures for $N_R = 4$ showing the limiting profiles. For a body with the same materials one subplot shows the case of the faster wearing material on the outside ( $k_2/k_1 = 100$ ) and the other one shows the case of the slower wearing material on the outside ( $k_2/k_1 = 0.01$ ). Obviously, the slope of the combination with the slower wearing material on the outside is much smaller because the outer ring 'limits' profile change in the inner rings. The result is a much smaller surface gradient. For the case of only two rings there are even some ratios resulting in a slightly lower surface gradient than that of a homogeneous cylinder.

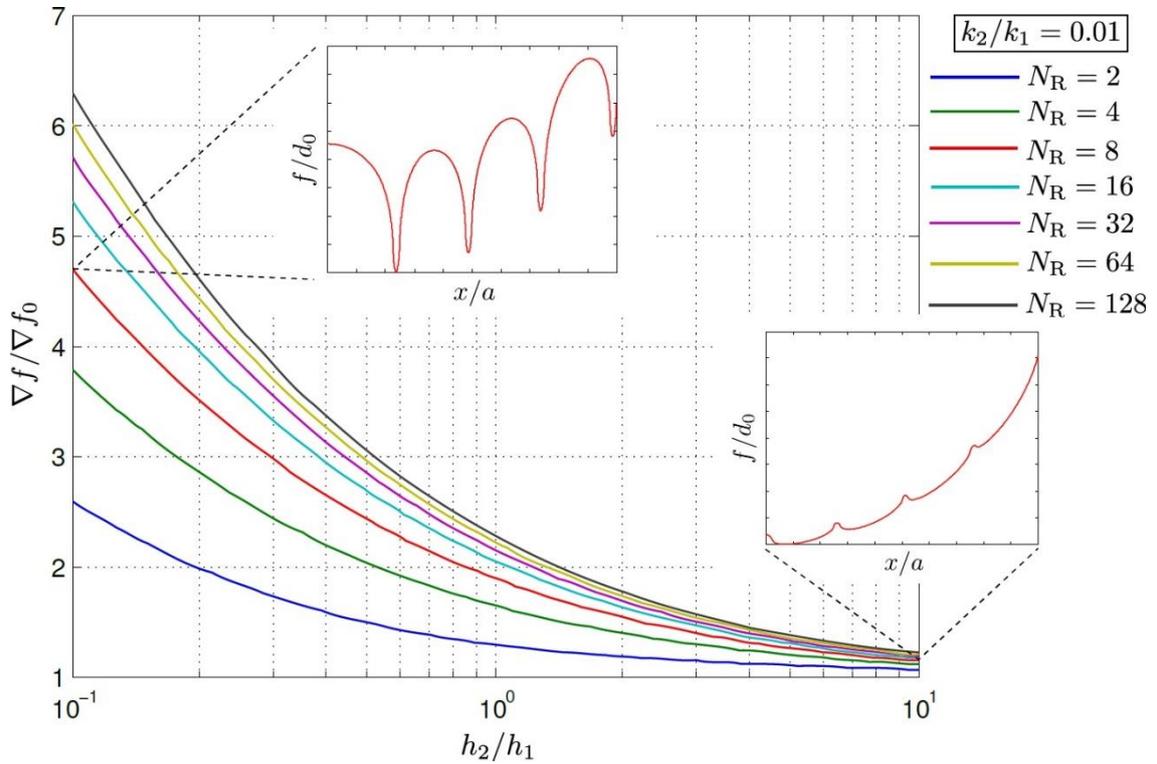

**Fig. 7** Normalized RMS of the surface gradient as a function of $h_2/h_1$ for different numbers of rings. The small figures show limiting profiles for extracted configurations

The variation of ring widths shown in Fig. 7 is done with a fixed ratio of wear coefficients $k_2/k_1 = 0.01$. Therefore, for $h_2/h_1 = 1$ the values for the different ring numbers are the same as in Fig. 6. For this configuration the outer ring is always consisting of the slower wearing material. If the ratio $h_2/h_1$ is increased, the share of slower wearing material is increased. Again, looking at the limiting profiles depicted for $N_R = 8$ it is obvious that for $h_2/h_1 = 10$ the parts of faster wearing materials are suppressed and have almost no influence on the profile. This means the surface gradient is very close to that of a homogeneous body. Thus, in all cases convergence towards $\nabla f/\nabla f_0 = 1$ can be observed.

However, if the share of slower wearing material is increased, the surface gradient is significantly higher. From the limiting profile for $N_R = 8$ and $h_2/h_1 = 0.1$ it is obvious that in these cases the problem is equivalent to only the thin rings with the number $N_R/2$ pressed into the elastic half-space. In this case there is no convergence and the surface gradient increases with the number of rings.

## 5. Conclusion

In this paper wear of an annular cylinder made of alternating rings of materials with different wear coefficients was analyzed. This was done by implementing two different methods. The Method of Dimensionality Reduction (MDR) was used to analyze time-dependent wear and to understand the influence of stress distribution and distribution of wear coefficients on the profile change. It was found that profile change slows down significantly as the limiting profile is approached. The limiting profile, that is eventually reached, is the state when the profile does not change anymore. In this state the stress distribution is piecewise constant.

As another approach to calculate the limiting profile, a method of direct integration was employed as well. Although the same limiting profile can be computed using both methods, this method of direct integration is much faster and was thus used to study the influence of the ratio of wear coefficients and ring widths on the limiting profile.

The root mean square of the surface gradient was the most important factor in this analysis as the friction coefficient is highly dependent on it. It was found that the surface gradient reaches a threshold for materials with very different wear coefficients. For coefficients with a difference of less than factor 10 the greatest dependence on the ratio of coefficients is observed. It was also found that for very few rings the order of rings is important and changes the surface gradient significantly.

The ratio of widths of the alternating rings influences the surface gradient strongly. If the width of rings of material with higher wear coefficient is very small, there is almost the same surface gradient as for a homogeneous cylinder, whereas the surface gradient is significantly raised when the width is very large.

### Acknowledgement
The authors thank V.L. Popov for valuable discussions.